\documentclass[a4paper,11pt]{article}
\pdfoutput=1 

\usepackage{jinstpub} 

\usepackage{amsmath}
\usepackage{graphicx}
\usepackage{subcaption}
\usepackage{url}
\usepackage{hyperref}

\title{\boldmath  Cerium-Doped Fused-Silica Fibers
as Wavelength Shifters}

\author[a,1]{N.~Akchurin,\note{Corresponding author.}}
\author[d]{N.~Bartosik,}
\author[a]{J.~Damgov,}
\author[a]{F.~De Guio,}
\author[c]{G.~Dissertori,}
\author[b]{E.~Kendir,}
\author[a]{S.~Kunori,}
\author[a]{T.~Mengke,}
\author[c]{F.~Nessi-Tedaldi,}
\author[d]{N. Pastrone,}
\author[c]{S.~Pigazzini,}
\author[b]{\c{S}.~Yaltkaya}

\affiliation[a]{Texas Tech University, Department of Physics and Astronomy,  Lubbock, TX, 79409, USA}
\affiliation[b]{Akdeniz University, Department of Physics, Antalya, 07070, Turkey}
\affiliation[c]{ETH, Z\"urich, Switzerland}
\affiliation[d]{INFN-Torino, Italy}
\emailAdd{nural.akchurin@ttu.edu}

\abstract{
We have evaluated the performance of a Ce-doped fused-silica fiber as wavelength shifter coupled to a CeF$_3$ crystal using electron beams at CERN. The pulse shape and collection efficiency were measured using irradiated (100~kGy) and un-irradiated fibers. In addition, we evaluated the light yield of various Ce-doped fibers and explored the possibility of using them in the future, including for precision timing applications in a high-luminosity collider environment.
}

\keywords{Cerium, optical fibers, fused silica}

\begin{document}
\maketitle
\flushbottom

\section{Introduction}
\label{sec:intro}
We have described our results concerning five prototype Ce-doped fibers in two previous publications:  in the first paper~\cite{JINSTPaper}, we concentrated on the scintillation pulse shape, light yield, attenuation length, and differences between the scintillation light and Cherenkov radiation.  In the second paper~\cite{JINSTPaper2}, we focused on fiber morphology, chemical composition, photo- and radio-luminescence, radiation-damage induced attenuation measurements and its modeling based on rate equations, and evaluation of changes in effective numerical aperture under gamma irradiation.   In each of our studies, we included clear fused-silica fibers containing no dopants  as the benchmark. In this paper, we present our results on the use of Ce-doped fibers, both irradiated and un-irradiated, as wavelength shifters when coupled to a single cerium fluoride (CeF$_3$) crystal using high-energy electron beams at CERN.  We also evaluated the performance of these fibers for use in precision timing detectors and compared them to common plastic scintillators and to a Cherenkov radiator (clear fused-silica fiber).  With these measurements, we were able to confirm and improve our previous light yield estimates.  All fibers reported on here were produced by Polymicro Technologies\footnote{Polymicro Technologies is a subsidiary of Molex located in Phoenix, AZ, USA}.

\section{Experimental Setup}
\label{sec:experimentalsetup}


The tests described here were performed in the H4 beam line of the CERN SPS North Area, which provides high-purity electron beams with energies between 20 and 250~GeV.  Four plastic scintillating counters of varying size, the smallest of which has a transverse dimension of $1 \times 1$ cm$^2$, were placed in the beamline for triggering purposes.  The data were acquired with a custom DAQ system~\cite{r-MAR}, which utilizes a CAEN\footnote{CAEN S.p.A., Via Vetraia 11, 55049 - Viareggio (LU) - Italy} V1742 VME board. The V1742 includes 4 DRS4 chips~\cite{r-DRS4} that allow sampling of the detector signal at frequencies up to 5 GHz. The choice of this flexible configuration enabled us to optimize the signal reconstruction offline as described in Section~\ref{sec:lightyield}.

\begin{figure}[ht]
\begin{center}\vspace{-1pc}
      \includegraphics[width=10 cm]{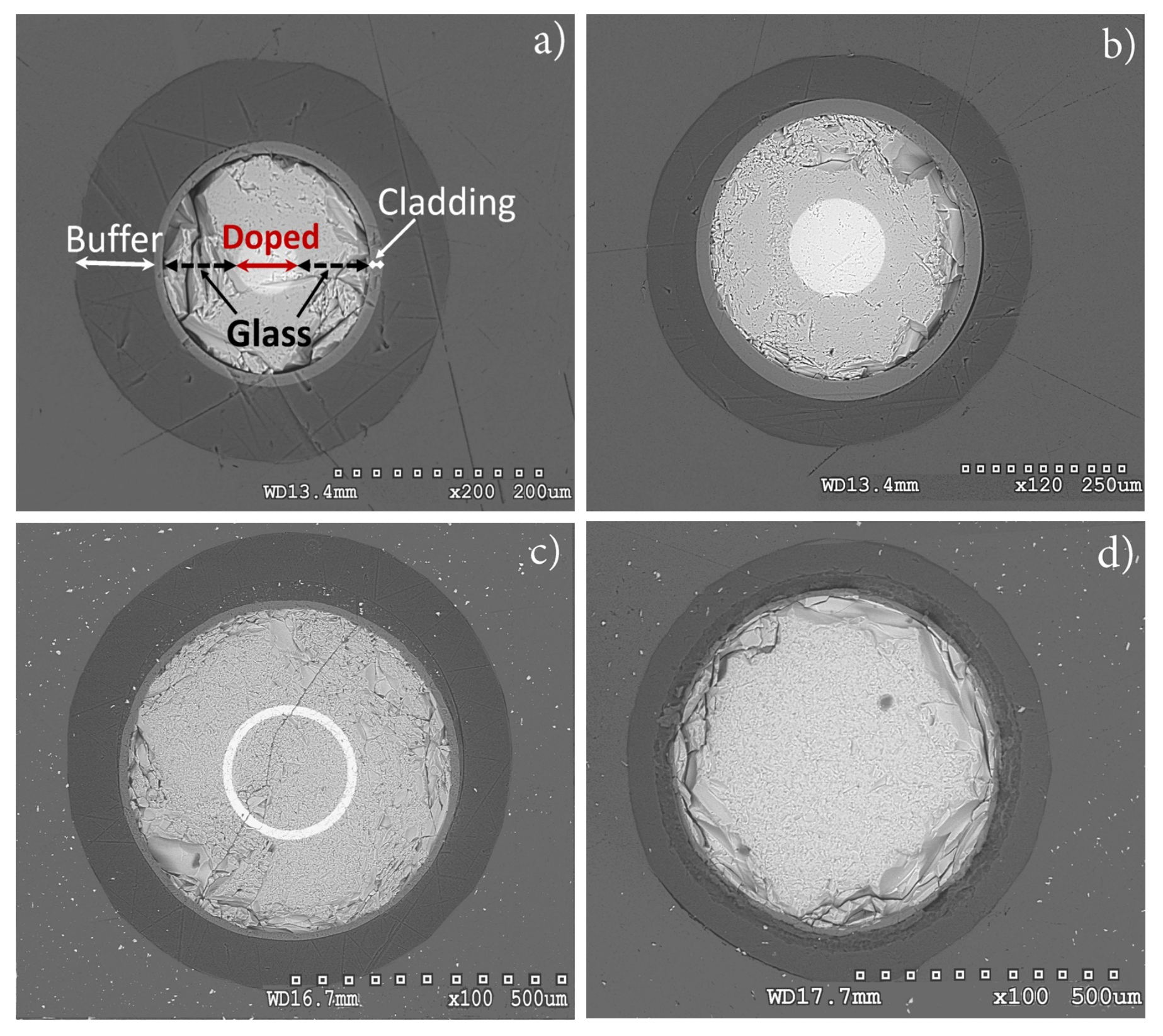}
\caption{\small The cross-section of fibers
{(a)} Phase-I,
{(b)} -II,
{(c)} -IV, and
{(d)} clear fused-silica are shown.  The terminology used throughout the paper for different parts of the fiber is indicated on panel {(a)}.}
    \label{fig:FiberCrossSections}
\end{center}
\end{figure}

Detailed information on the fibers can be found in~\cite{JINSTPaper}, and Table~\ref{tab:fiberlist} summarizes their physical dimensions.  Figure~\ref{fig:FiberCrossSections} displays cross-sectional scanning electron microscope (SEM) images obtained by the Hitachi-S-4300~SE/N.  Some performance characteristics of Phase-I and -II fibers that were measured with high energy beams at Fermilab were reported in our first paper \cite{JINSTPaper}.  For example, the Phase-I optical attenuation length was measured as $5.8\pm0.7$ m using 16~GeV  electrons.  The propagation speed of light was $19.50\pm0.38$ cm/ns and was found to be consistent with previously reported values \cite{Goro,Akch97}.  When the tail of the pulse was modeled by a sum of two exponential functions, two decay constants were obtained, $\tau_1= 20.8\pm5.4$ ns and $\tau_2= 93.0\pm12.6$ ns.  The contributions of these components to the total radio-luminescent light emission turned out to be $23.9\pm5.3$\% and $76.1\pm5.3$\%, respectively.  In our second paper \cite{JINSTPaper2}, we focused on the fiber chemical compositions and their impact on light emission and radiation-hardness properties.  For example, the presence of aluminum, which is often introduced to prevent dopant clusterization in glasses, shifts the emission spectrum by 10-20~nm to higher wavelengths.  We also identified two strong photo-luminescence lines peaking at $\sim$420 and $\sim$460~nm due to the transition from the lowest 5$d$ to the 4$f$ of the Ce$^{3+}$ levels when the spin-orbit coupling splits the transition into two ($^2F_{J}=7/2$ and $5/2$) separated by $\sim$2200 cm$^{-1}$. The most radiation-hard fiber, the Phase-IV type, with a coaxial Ce-doped ring structure, maintains a fraction of its photo-luminescence ($\sim$20\% of the original value) and transmission (4~dB/m) in the 430 to 600~nm range after 100~kGy of accumulated radiation. Although there is room for improvement, this performance represents significant progress over our previous attempts towards engineering a radiation-hard inorganic scintillating  fiber.  We also showed that radiation induced attenuation and recovery can be successfully modeled by a set of self-consistent kinematic rate equations.

\begin{table}[htp]
\caption{\small Listed are the investigated fibers with their dimensional characteristics. In particular the outside diameter (OD) is reported for the fiber's core, glass, cladding and buffer. All fibers were cladded with fluorinated acrylate and UV-cured acrylate buffer.}
\vspace{-2 mm}
\begin{center}
{\small
\begin{tabular}{|l|c|c|c|c|c|c|c|c|} \hline
Fiber              &  Core OD            & Glass OD        & Clad OD          & Buffer  OD        \\
Name               &[${\rm \mu m}$]      & [${\rm \mu m}$] & [${\rm \mu m}$]  & [${\rm \mu m}$]   \\
\hline
                    &                     &                &                  & \\
Phase-I            & 60$\pm$7            &	200$\pm$6      & 230$^{+5}_{-10}$ &	350$\pm$15        \\
Phase-II           & 150$\pm$20		     &	400$\pm$10     & 430$\pm$10	      &	550$\pm$30        \\
Phase-IV           & 200$\pm$8(clear)   &	600$\pm$6	   & 630$\pm$10       &	800$\pm$30        \\
                          & 230$\pm$8(doped) &	   &     &	         \\
Clear fused-silica & 600$\pm$10          &-				   & 630$^{+5}_{-10}$ &	800$\pm$30	      \\
                    &                     &                &                  & \\
\hline
\end{tabular}
}
\end{center}
\label{tab:fiberlist}
\end{table}

\section{Scintillation Light Yield Measurements}
\label{sec:lightyield}
\begin{figure}[ht]
\begin{center}\vspace{-1pc}
      \includegraphics[width=12cm]{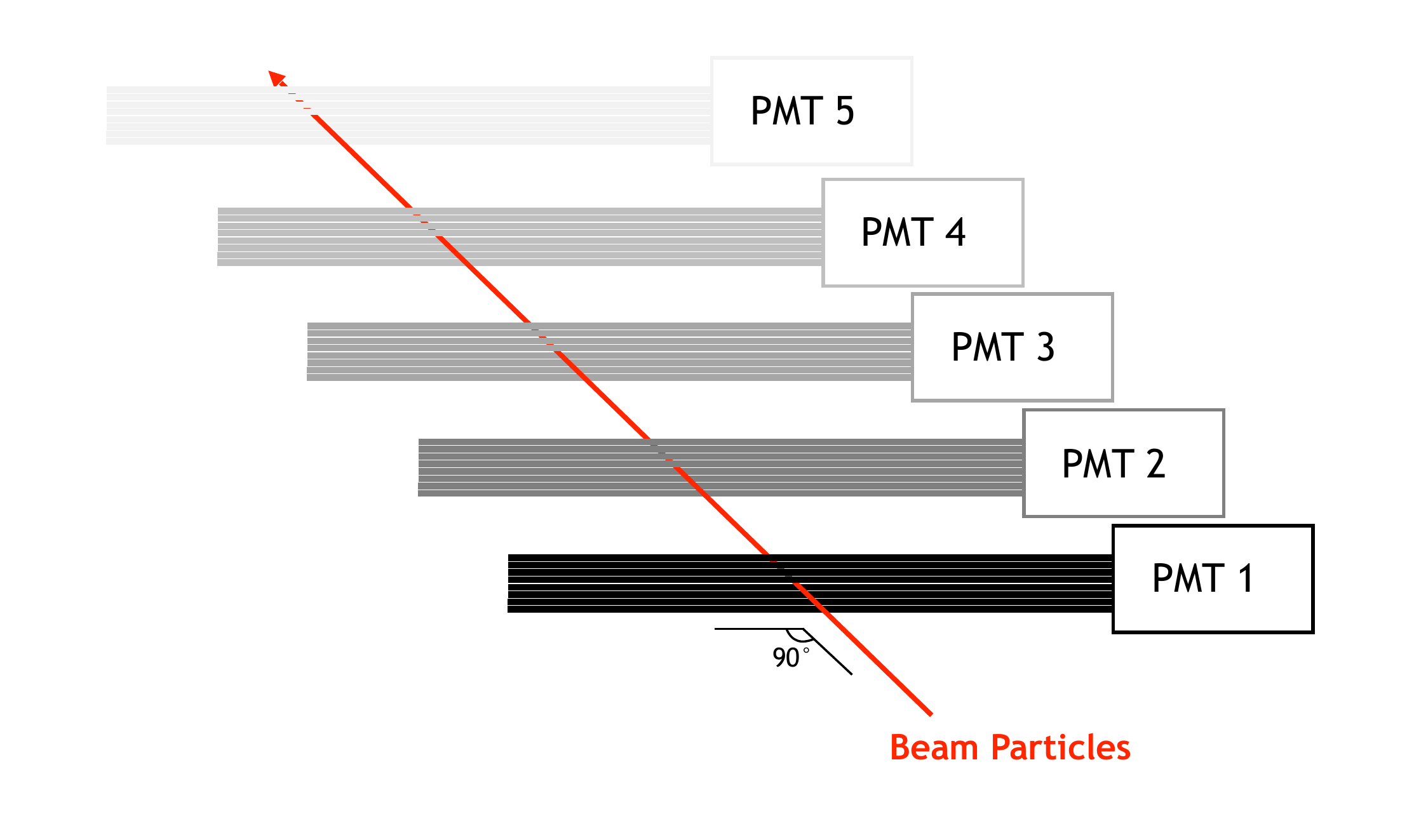}
\caption{\small The fiber bundles, about 40 cm long, were positioned parallel to each other and oriented at 90$^\circ$ with respect to the beam line. The light was detected with a PMT at one end of each bundle.}
    \label{fig:bundles}
\end{center}
\end{figure}

Light yield is an important characteristic of the fibers, and we adopted the same method to evaluate it as employed in the first paper~\cite{JINSTPaper}.   The fibers were cleaved, bundled, inserted in thin aluminum tubes, and exposed to a 25~GeV electron beam whose direction was at $90^{\circ}$ with respect to the axis of the fiber bundles (Figure~\ref{fig:bundles}). A copper block, 4~cm (2.7~$X_0$) thick, was placed in front of the bundles to generate multiple particles that enhance light generation in the fibers. The light was collected by Hamamatsu R6427\footnote{HAMAMATSU Photonics, Electron Tube Division, Shimokanzo, Iwata City, Japan} photomultiplier tubes (PMT). The PMT signals were sampled at a frequency of 2.5~GHz, and a common trigger was generated by a fast Microchannel Plate (MCP)~\cite{r-MCP} located in front of the copper block in the beam line. Each bundle contained only one fiber type: 6 Phase-IV un-irradiated, 6 Phase-IV irradiated, 85 Phase-II, or 61 clear fused-silica fibers. The absolute light yield was measured in photo-electrons (p.e.) per event. The calibration was determined by the measurement of the single photo-electron peak for each PMT. Correction due to the difference in the quantum efficiency (QE) of the PMTs is not applied in this measurement. The relative deviation of the QE is know to be within 15\% of its nominal value . We used the timing characteristics of the Cherenkov ($Q$) and scintillation ($S$) signals to separate and integrate their contributions to the total signal individually. We measured $Q$~=~0.092 p.e. and $S$~=~0.147 p.e. for the un-irradiated Phase-IV fiber bundle per event. For the irradiated Phase-IV fibers, we obtained $Q$~=~0.058 p.e. and $S$~=~0.62 p.e.  These data clearly indicate that the Cherenkov light yield was reduced after irradiation. The measured scintillation light yield however increased by about a factor of four. We also measured $Q$~=~0.44 p.e. and $S$~=~2.25 p.e. per event for the Phase-II bundle.

We used the GEANT4 simulation package~\cite{r-GEANT4} to translate these measurements into a characteristic scintillation light yield for the Ce-doped fibers.  We implemented the full geometry of the setup (descriptions of the copper block and individual fiber structures, number of fibers in a bundle, and positions of the bundles), a detailed description of light propagation (ray tracing within and at the exit of each fiber), and the characteristics of the photodetectors (quantum efficiency as a function of wavelength, transit time spread, and rise time).  We compared the measured $S/Q$ ratio with the GEANT4 prediction.  This approach (using the $S/Q$ ratio) resulted in the cancellation of many systematic uncertainties, such as the beam and bundle alignment, optical coupling of the bundles to PMTs, and uncertainty in the single photo-electron signal definition in the measurement.  The scintillation mechanism in GEANT4 is implemented such that 500 photons are produced isotropically per MeV of energy deposited via ionization losses from charged particles in the Ce-doped fused-silica. The wavelength of these photons was  matched to the measured emission spectrum of the fibers (excited by 337 nm $N_2$ laser light as described in \cite{JINSTPaper2}).  As a result, the measured characteristic scintillation light yield for the Ce-doped fused-silica was found to be $1130\pm 500$ photons/MeV for the un-irradiated Phase-IV fibers. The dominant systematic error arises from uncertainties in the inner and outer radii of the Ce-doped ring. 

\begin{figure}[ht]
\begin{center}
      \includegraphics[width=4.8cm]{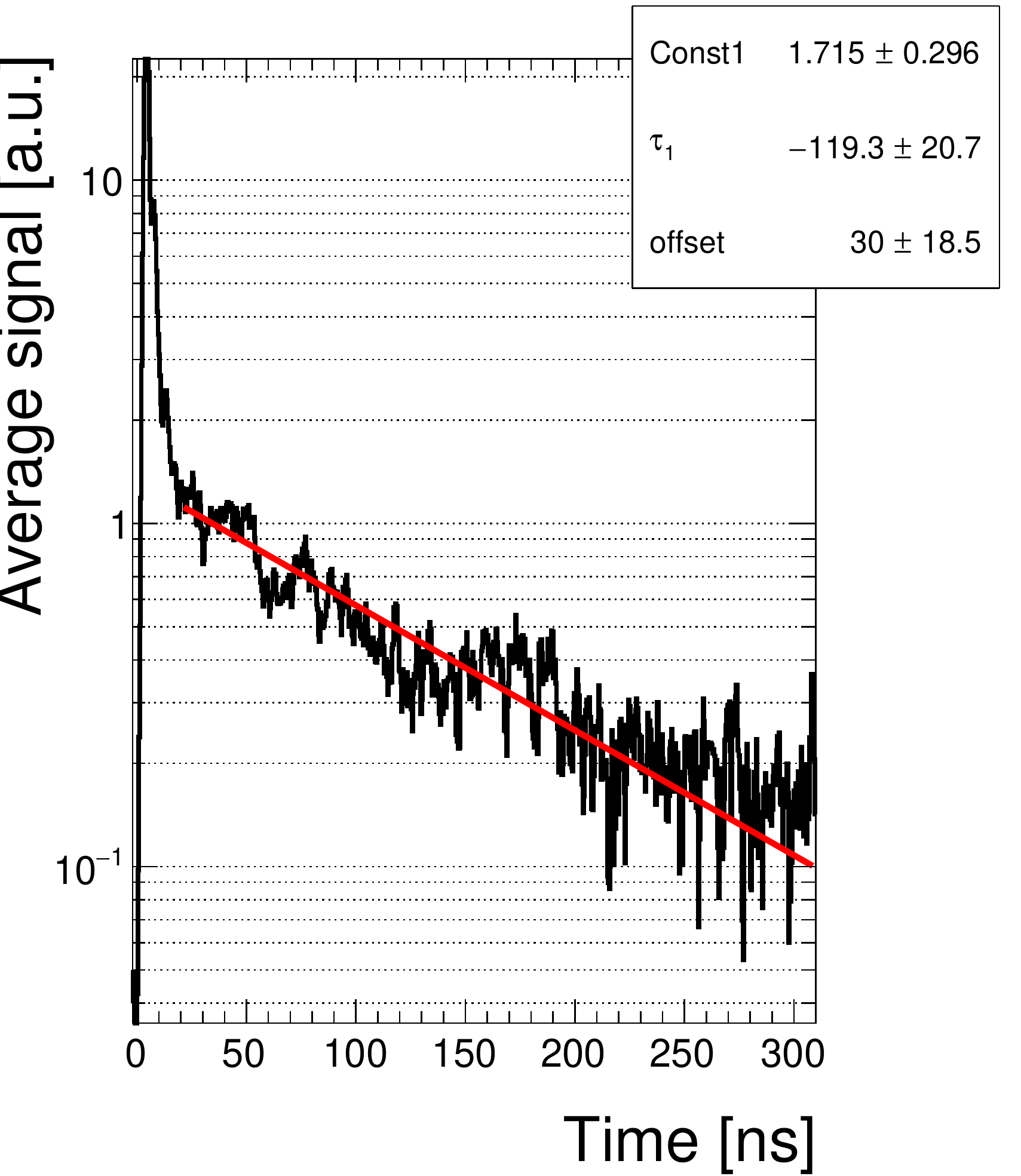}
      \includegraphics[width=4.8cm]{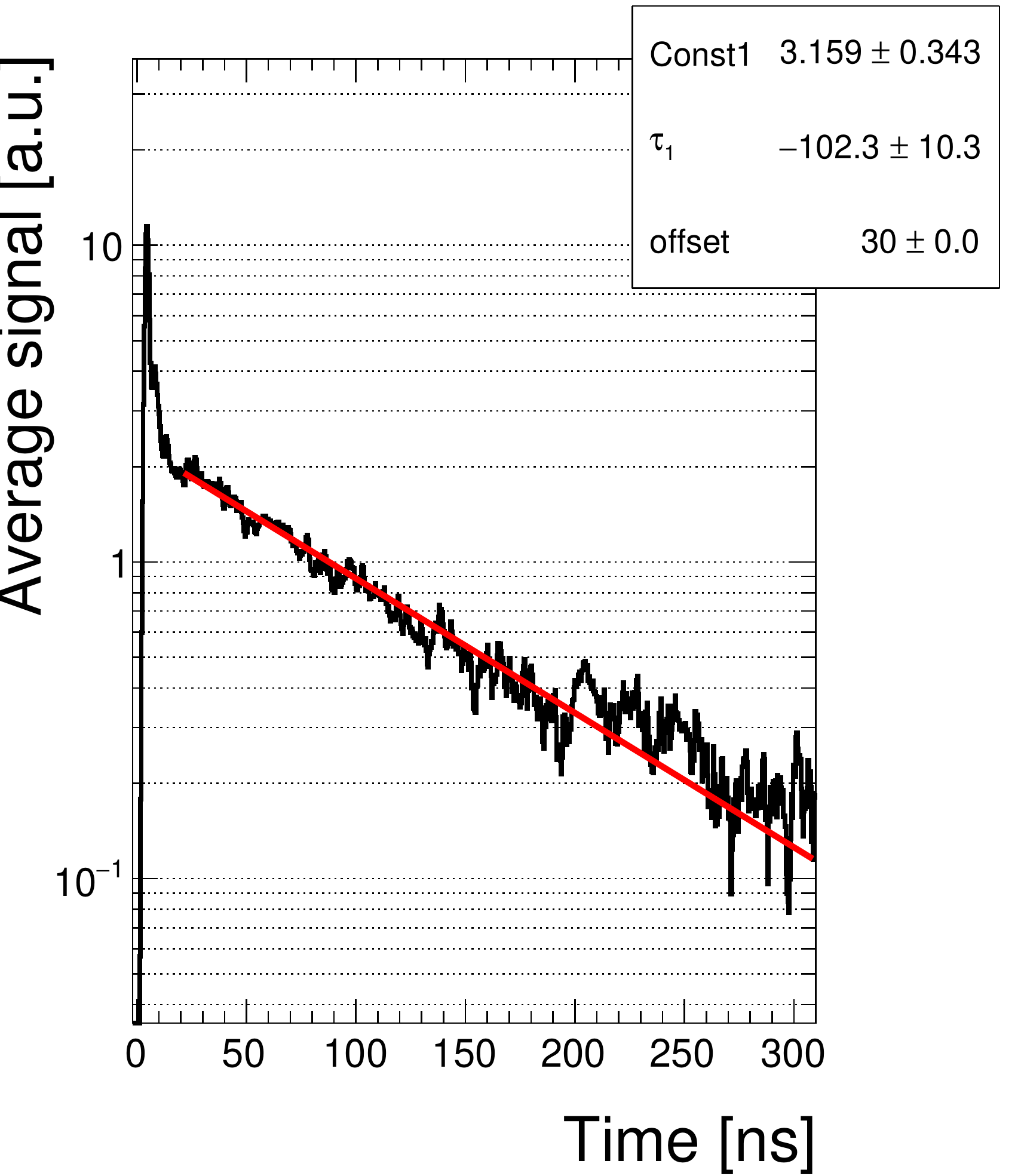}
      \includegraphics[width=4.8cm]{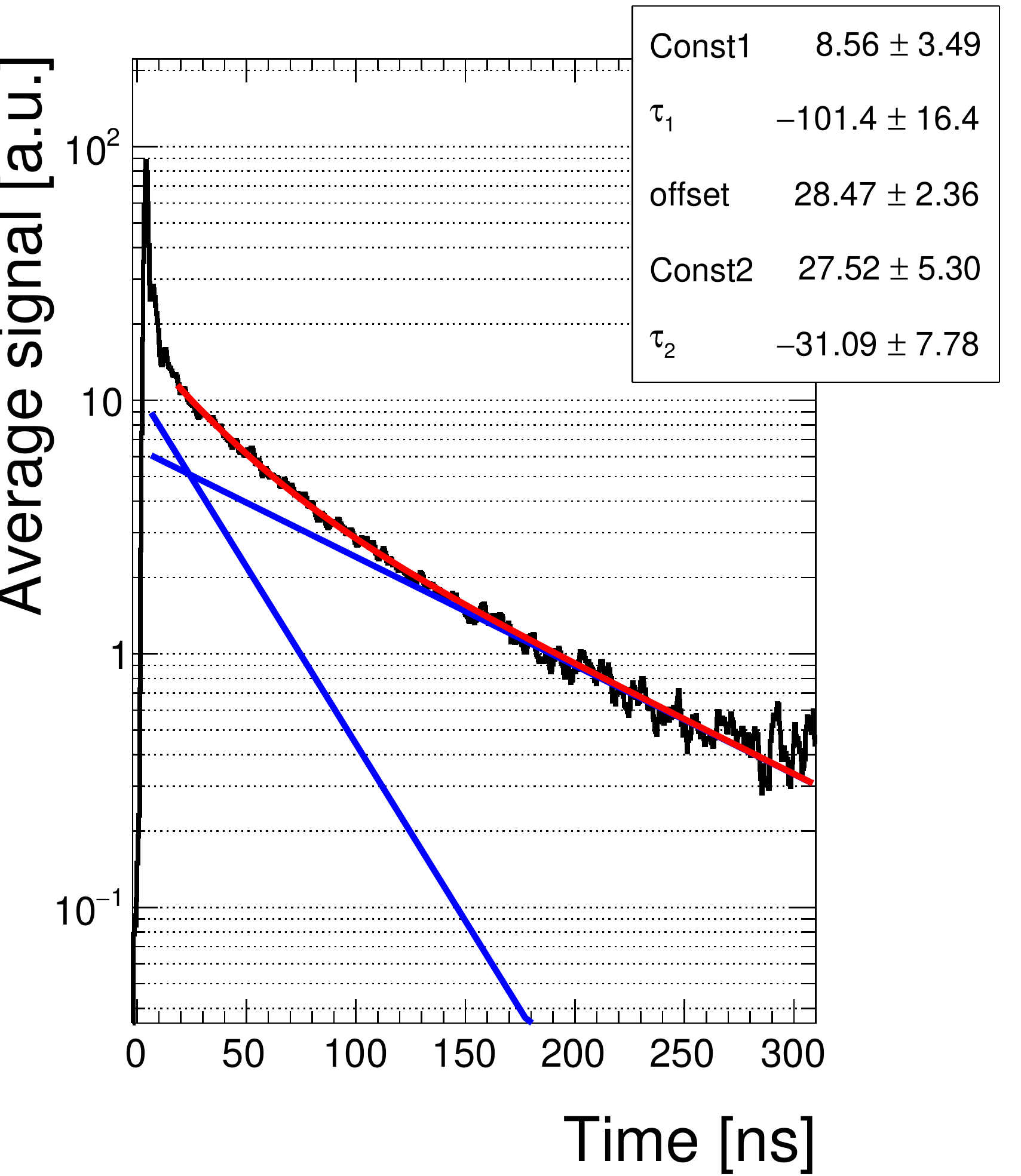}
\caption{\small The average signal from the un-irradiated Phase-IV (left), irradiated Phase-IV (middle), and Phase-II (right) fibers: The tail of the scintillation signal is well described by a single exponential (red line), with a decay constant of $119.3 \pm 20.7$ ns and $102.3 \pm 10.3$ ns for the Phase-IV un-irradiated and irradiated fibers, respectively. The tail of the Phase-II scintillation signal is better described by a sum of two exponentials, with decay constants of $101.4 \pm 16.4$ ns and $31.1  \pm 7.8$ ns and relative contributions of $66 \pm 16 \%$ and $34 \pm 16 \%$.}
    \label{fig:phase4scintTime}
\end{center}
\end{figure}

We also measured the timing characteristics of the scintillation component of the signals for the Phase-IV fiber. We found that a single exponential function describes the tail best (Figure~\ref{fig:phase4scintTime}). The decay constants were $119.3 \pm 20.7$ ns and $102.3 \pm 10.3$ ns for the un-irradiated and irradiated fibers, respectively which are found to be consistent within their uncertainties. This allows to state that the irradiation did not alter the scintillation decay constant value. A sum of two exponential functions was used to model the Phase-II signal. The decay constants were $101.4 \pm 16.4$ ns and $31.1  \pm 7.8$ ns, with relative contributions of $66 \pm 16 \%$ and $34 \pm 16 \%$ to the signal. We tested the double-exponential model on the un-irradiated and irradiated Phase-IV signals with a constrained decay constant of $31$ ns for the fast component: we found that the fit favored the relative contribution of $10 \pm 20\%$ and $6 \pm 11\%$, respectively. This is consistent with the absence of a fast component with a decay constant of $31$ ns.

\section{Ce-doped Fibers Coupled to a CeF\texorpdfstring{$_3$}{} Crystal}
\label{sec:WLS}
\begin{figure}[ht]
\begin{center}
      \includegraphics[width=10cm]{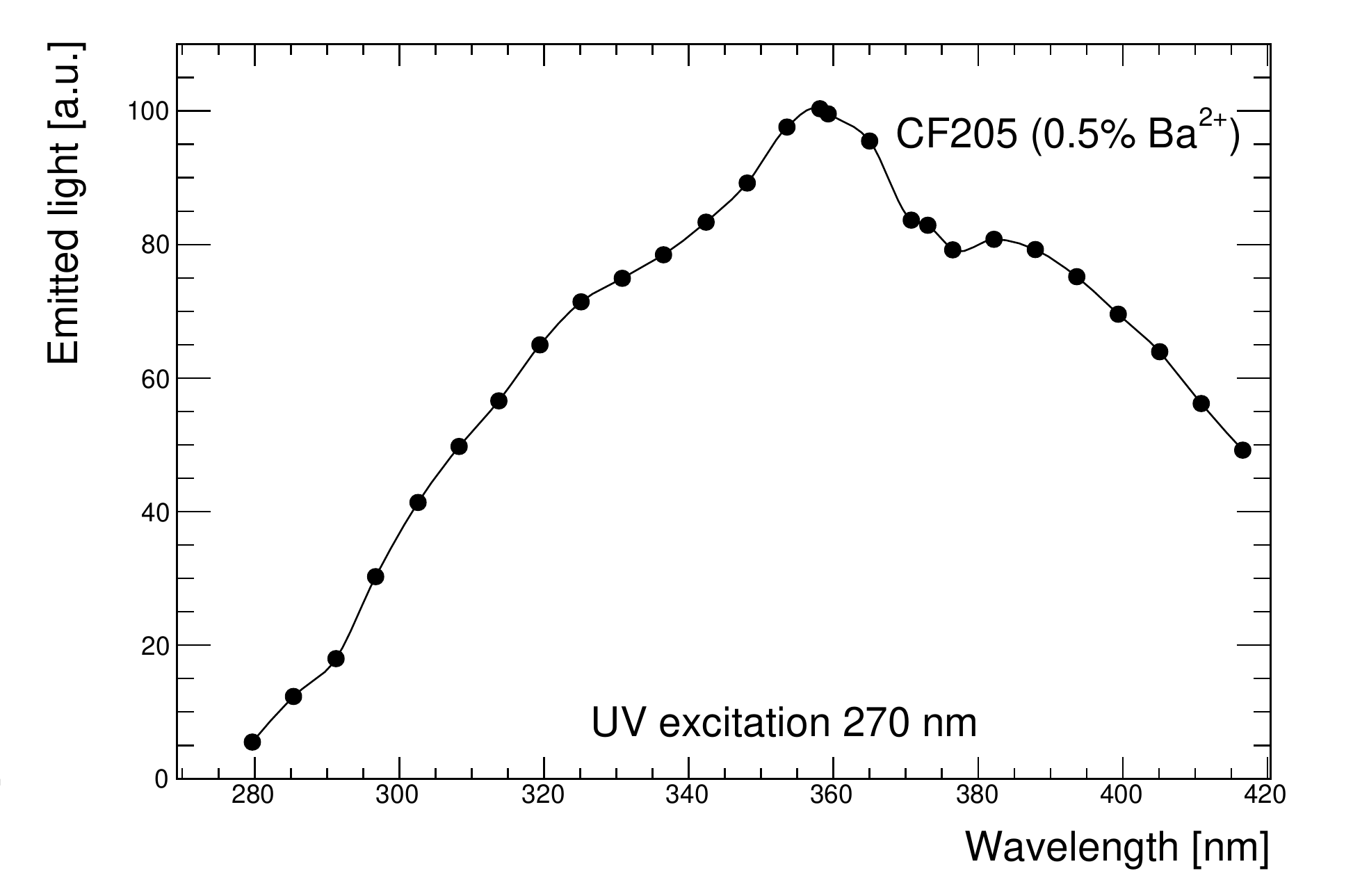}
\caption{\small Photo-luminescence spectrum for a barium-doped CeF$_3$ crystal~\cite{r-EACEF3}.}
    \label{fig:cef3spectrum}
\end{center}
\end{figure}

Cerium fluoride is an intrinsic scintillator with a density $\rho=6.16\; {\mathrm{g/cm}}^3$, radiation length $X_{\rm o}~=~1.68$~cm, Moli\`ere radius $R_{\rm M}~=~2.6$~cm, nuclear interaction length $\lambda_{\rm I}~=~25.9$~cm, and refractive index $n = 1.68$, whose luminescence was discovered by F.~A.~Kr\"oger and J.~Bakker~\cite{r-KRO} in 1941. In the 1990s, cerium floride was extensively studied~\cite{r-AND,r-MOS} as a bright and relatively fast scintillator, with a decay time constant of 10 to 30~ns. Cerium fluoride was considered for electromagnetic calorimetry for the Compact Muon Solenoid (CMS) experiment at the high-luminosity Large Hadron Collider (HL-LHC) because it recovers from radiation damage~\cite{r-NIMCEF3}, unlike many other crystals~\cite{r-FISSNIM, r-NIMLYSO}. A sampling calorimeter design was proposed, alternating tungsten as absorber and cerium fluoride as scintillating medium, with a readout through wavelength-shifting cerium-doped quartz fibers along the chamfers of each calorimeter cell~\cite{r-CALORCEF3, r-WCEF3FRA, r-WCEF3H4}. This combination of crystal and fiber was chosen because cerium fluoride broadly emits between 300 and 400~nm (Figure~\ref{fig:cef3spectrum}), with spectrum shape variations depending on the presence of dopants~\cite{r-EACEF3}, conveniently in the domain of wavelengths that excite the photo-luminescent emission of cerium-doped fused-silica~\cite{r-vedda}.

\begin{figure}[ht]
\begin{center}\vspace{-1pc}
      \includegraphics[width=14 cm]{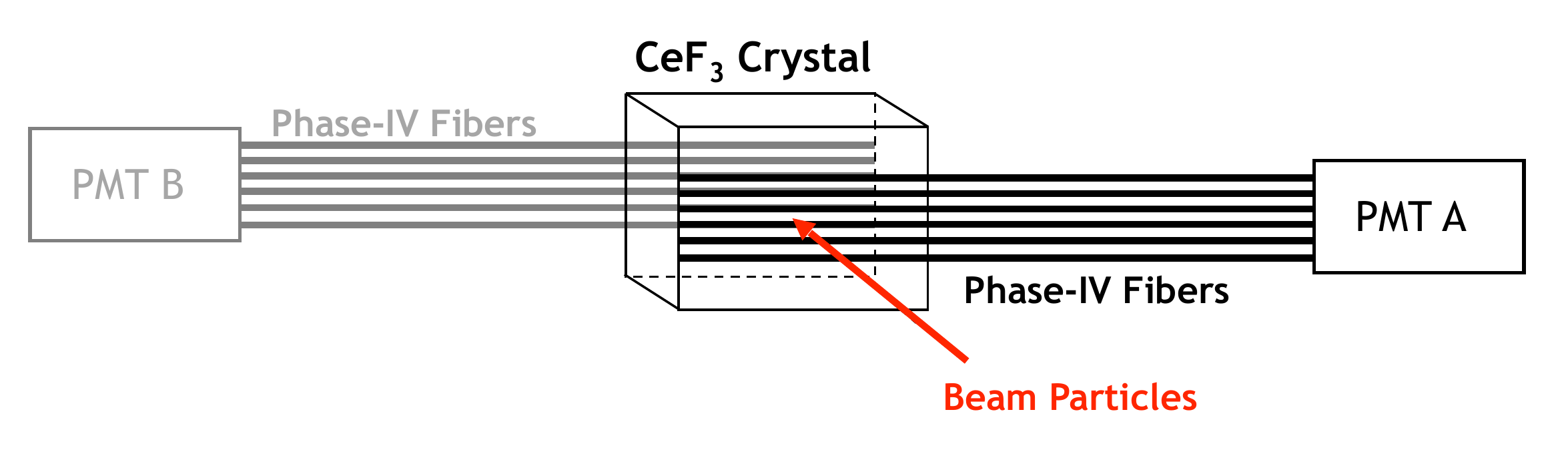}
\caption{\small Arrangement of the fibers with respect to the CeF$_3$ crystal.}
    \label{fig:CeF3Arrangement}
\end{center}
\end{figure}

For the test presented here, a small block of cerium fluoride crystal was used, of dimensions $33\times 24 \times13$ mm$^3$. All its surfaces were unpolished and diffusive to maximize the scintillation light escaping the scintillator and entering the fibers. White, diffusive Tyvek\texttrademark~(a product of DuPont\footnote{DuPont, Wilmington, DE, USA}) was wrapped around the four larger faces while the two small faces were left unwrapped. As shown in Figure~\ref{fig:CeF3Arrangement}, two sets of six fibers were air-coupled to the larger one of the crystal surfaces without using any optical couplant and held in place by a tightly wrapped Tyvek sheet. One set of six consisted of the un-irradiated Phase-IV fibers and the other one of irradiated ones that were viewed independently by different PMTs. The equalization of the signals from the two channels was performed offline and consisted of calibrating the gain of the two PMTs by equalizing the position of the single photo-electron peak. In addition to the PMT gain calibration, other sources of uncertainty included coupling of the fibers to the PMT window and possibly a different amount of light seen by the bundles when placed upstream or downstream of the crystal with respect to the beam direction. For calibration of the two channels, the PMTs and the orientation of fibers on the crystal surfaces with respect to the beam were swapped in all four possible combinations and equalized offline.

\begin{figure}[ht]
\begin{center}
        \includegraphics[width=7cm]{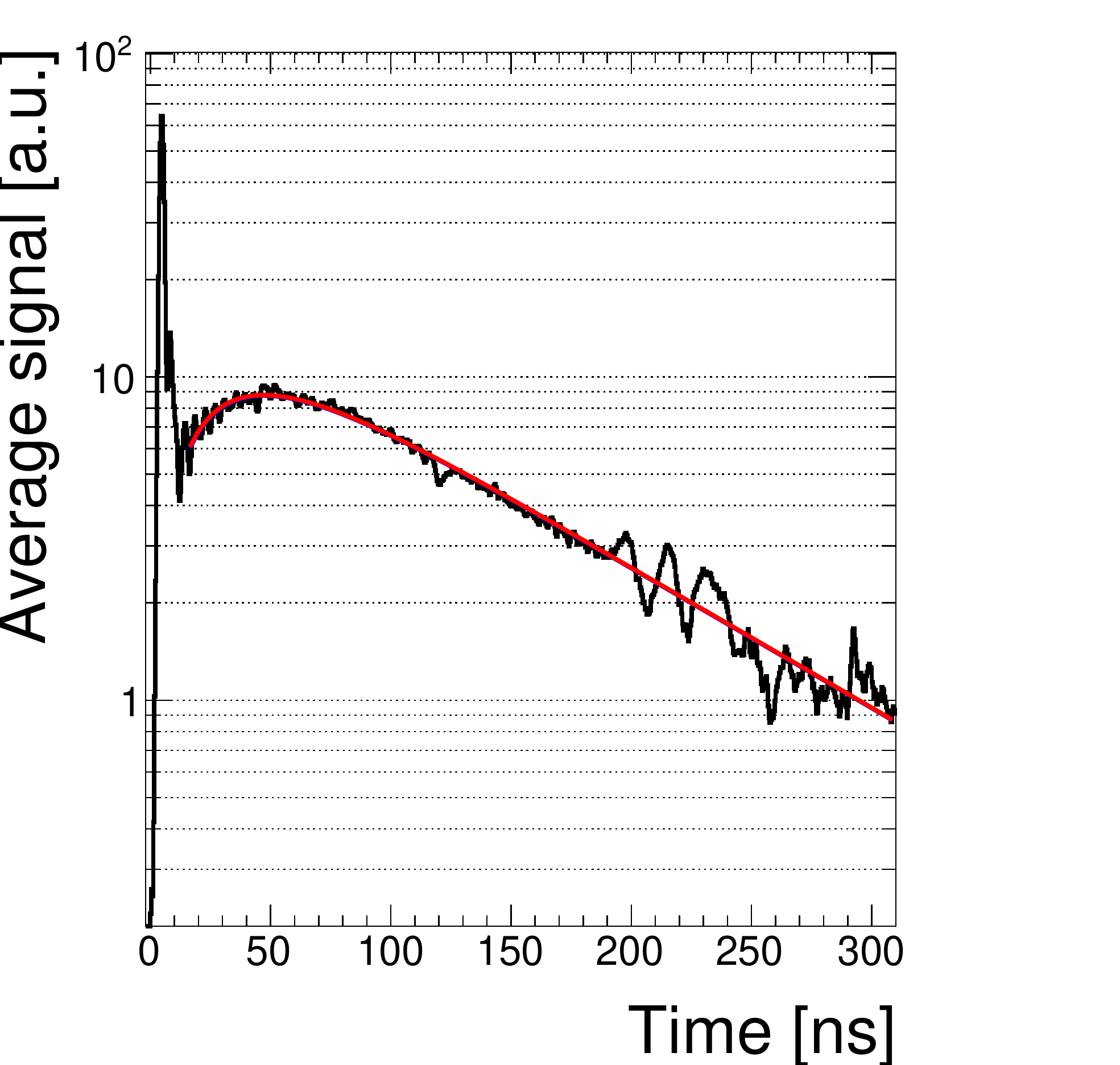}
        \includegraphics[width=7cm]{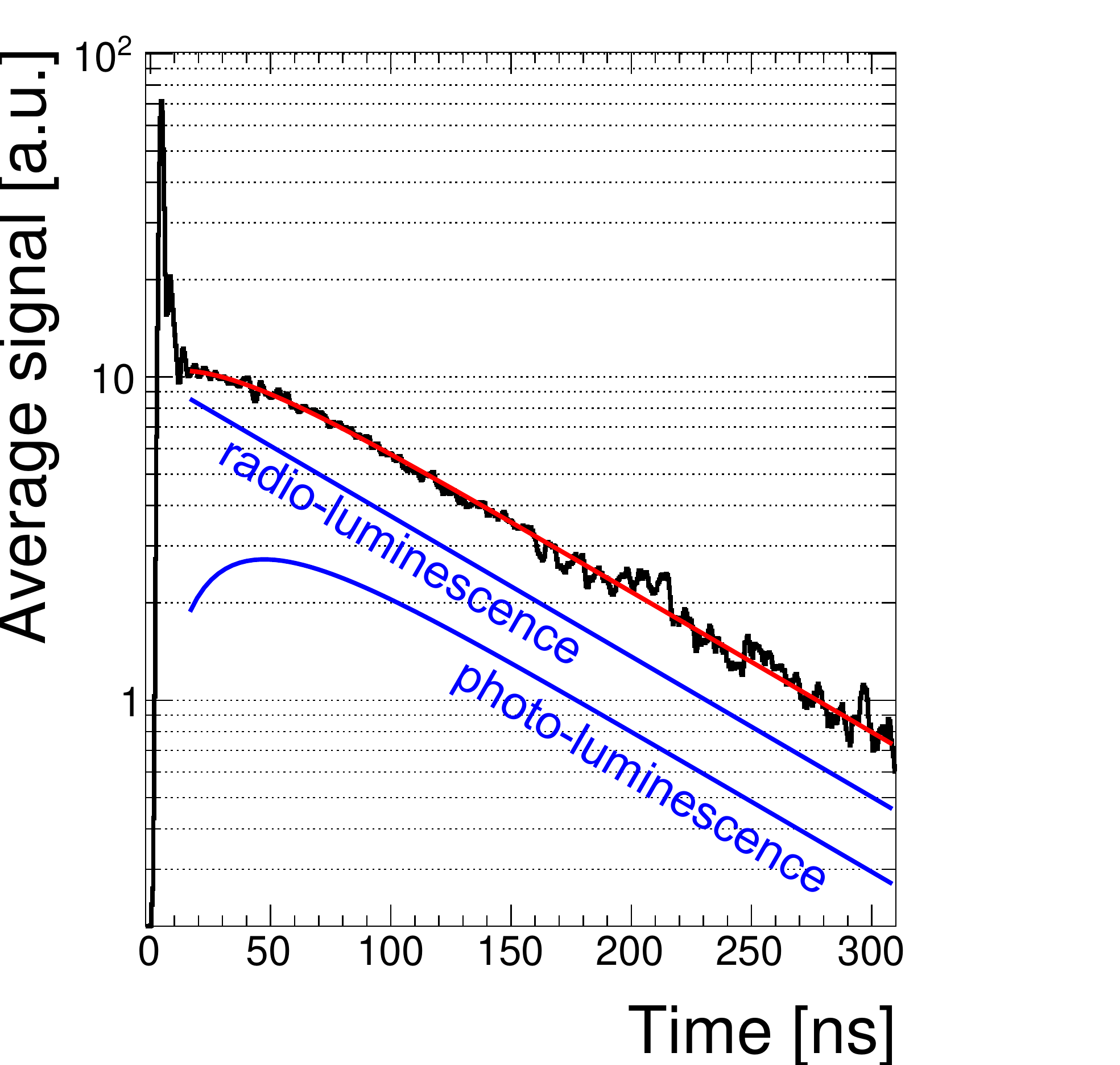}
    \caption{\small The averaged signals of wavelength-shifted light from un-irradiated (left) and irradiated (right) Phase-IV fibers coupled to the CeF$_3$ crystal. Blue lines show the radio- and photo-luminescence contributions to the signal. }
    \label{fig:CeF3signals}
\end{center}
\end{figure}

The light detected by the PMTs has two components: photo-luminescence and radio-lumine\-scen\-ce. Photo-luminescence causes the cerium fluoride scintillation light to be shifted in wavelength by the Ce-doped fibers, while radio-lumines\-cen\-ce results from charged particles traversing the Ce-doped ring in the fibers. The timing characteristics of the two components differ due to the presence of a comparable scintillation decay time ($\sim$26~ns) from the CeF$_3$ crystal in the photo-luminescence component. We used analytic functions to model the two contributions to the observed signal, as shown by blue lines in Figure~\ref{fig:CeF3signals}. We fit this model to the data in order to extract the relative contributions in the signal. The un-irradiated Phase-IV fiber signal shape is best described by a model with a $\sim100$\% contribution from the photo-luminescence, while the data for the irradiated Phase-IV fiber suggest that only a quarter of the total signal comes from photo-luminescence. This corresponds to a reduction in the light yield from photo-luminescence to a level of $28 \pm 5$\% of the original value as a result of the irradiation of the Phase-IV fiber. The outcome is not necessarily due to a reduction in photo-luminescent light yield from irradiation, it could also be due to a reduced transmission of the scintillation light from cerium fluoride through the cladding of the fiber.

\section{Timing Measurements}
\label{sec:timing}
In addition to the light-yield studies presented in the previous section, an attempt to characterize the timing performance of these fibers was evaluated because the Cherenkov component is naturally fast and well separated from the scintillation one. Studies indicate that precise timing information (tens of ps) is beneficial for pile-up mitigation and for the correct assignment of physics objects to their primary vertices at the HL-LHC~\cite{r-timingATLAS,r-timingCMS}. It then becomes worthwhile to exploit the Cherenkov signal for timing applications.  The relative abundance of the light detected due to the Cherenkov process  strongly depends on the orientation of the fibers with respect to the nominal direction of the shower axis (Figure~\ref{fig:bundles}). 

\begin{figure}[ht]
\begin{center}
      \includegraphics[width=12cm]{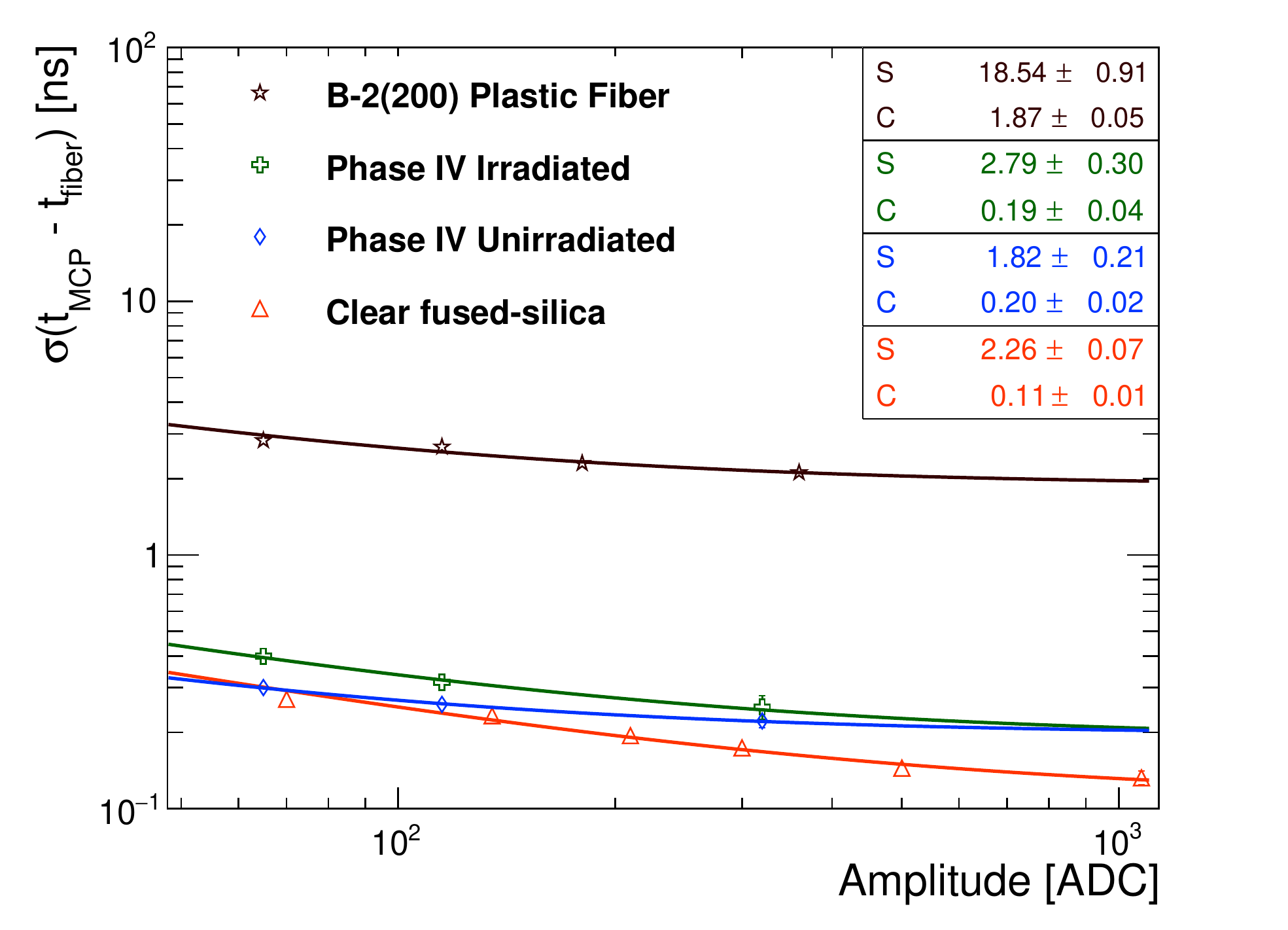}
\caption{\small Time resolution curves for different types of fibers (Eq~\ref{eq:eq}) where an MCP was used as a time reference, {\it i.e.} $(t_{\rm MCP} - t_{\rm fiber})$.  One ADC count equals 0.24~mV.}
    \label{fig:fibres_time_res}
\end{center}
\end{figure}

The fixed time reference was defined offline when the MCP pulse reached half of its maximum amplitude, mimicking a constant fraction discriminator at 50\%. The timing performance was then measured by fitting the distributions of $(t_{\rm MCP} - t_{\rm fiber})$ in bins of charge with a gaussian distribution.  The time resolution can be expressed as a sum of two terms that account for different sources of uncertainty in quadrature and parametrized as follows:
\begin{equation}
    \sigma_t^2 = \bigg( \frac{s}{\sqrt{A}} \bigg)^2 + c^2
    \label{eq:eq}
\end{equation}
where $A$ is the measured amplitude while $s$ and $c$ are the stochastic and constant terms, respectively. For the MCP, the measured constant term is $17.0\pm0.3$~ps~\cite{r-MCP}.  The time resolutions for the irradiated and un-irradiated Phase-IV and clear fused-silica fiber were measured and compared to the performance of a B-2(200) plastic scintillating fiber~\cite{r-B2} by Kuraray\footnote{Kuraray Co., Ltd., Chiyoda, Tokyo, Japan} (Figure~\ref{fig:fibres_time_res}).  The following observations can be made:

\begin{itemize}
\item There is a clear advantage in using Cherenkov light for timing measurements compared to the slower scintillation light. For the setup considered, where very few photo-electrons were generated per event, promptly produced photons guarantee a time resolution that is not degraded by the slower light re-emission from scintillation processes. In this case, the scintillating plastic fiber utilized had a characteristic decay time constant of $\sim$5~ns.
\item In the hypothesis of an infinite number of photo-electrons, the cerium-doped fused-silica fibers and the scintillating plastic fibers are expected to have the same timing performance if they were to be read out by the same photodetector. Our measurements, however, did not allow a precise evaluation of the constant terms due to a relatively low signal.
\item The inferior timing performance of the Phase-IV fiber with respect to the clear fiber is the direct consequence of the higher photo-statistics available in the latter. The number of Cherenkov photons produced by a particle traversing the bundle is proportional to the path length of the particle in the fiber. As described in Sections~\ref{sec:experimentalsetup} and \ref{sec:lightyield}, the effective section of the fused-silica fibers bundle is larger compared to the section of the fibers in the Phase-IV bundles.
\item A degradation of timing performance is visible for the irradiated fibers compared to the un-irradiated ones. This observation is interpreted as a worsening of the light transmission due to the formation of new color centers in the fiber. Also, a possible systematic error associated with an imperfect coupling of the fibers to the PMT window can not be excluded.
\item The PMTs used in these studies were not optimized in terms of timing performance and thus degraded the timing measurements because of slow rise time (1.7 ns) and significant transit time spread (500 ps FWHM according to the datasheet~\cite{r-PMT}).
\end{itemize}
In conclusion, these data reaffirm that time resolution improves when prompt (Cherenkov-like) and abundant (scintillator-like) light is detected with a fast photosensor.

\section{Discussion and Conclusions}
\label{sec:conclusions}
Wavelength shifting fibers find their application in many fields of physics. At the colliders they have to be radiation-tolerant and fused-silica fibers are considered for this reason. In this paper, the performance of irradiated (100~kGy) and un-irradiated Ce-doped fused-silica fibers are compared when coupled to a CeF$_3$ crystal and exposed to an electron beam at CERN. In the signal modeling the photo- and radio-luminescence components are considered and the data analysis suggests that, while for the un-irradiated fibers bundle the signal comes entirely from photo-luminescence processes, for the irradiated case that fraction is reduced to one fourth while the importance of the radio-luminescence component increases. This observation seems consistent with the indications from the light yield measurements performed with the fibers bundles exposed directly to the beam where a sizable radio-luminescence contribution to the signal is seen for the irradiated fibers.

An interesting feature of the fibers considered is the presence of a fast signal component that originates from from Cherenkov processes. The Cherenkov peak is well separated from the scintillation one and could be exploited to obtain precise timing information, relevant for applications at colliders in environments with high particle multiplicity. Despite the fact that the readout used is not optimal for timing measurements, it was possible to measure the time resolution for the fused-silica fibers and compare it with the one for a plastic scintillating fiber that doesn't exhibit a Cherenkov peak. As expected, a better timing performance is achieved when promptly produced photons from Cherenkov processes are considered instead of photons from delayed scintillation processes.

In addition to timing applications, the presence of both the Cherenkov and the scintillation component is interesting for a dual readout calorimetry approach where the measurement of the electromagnetic fraction of the showers on an event-by-event basis would allow to improve the calorimeter resolution by compensating for the effects arising from large fluctuations of electromagnetic showers. While in the existing prototypes the readout for the Cherenkov and scintillation light is decoupled (clear fibers and scintillating fibers are alternated in the calorimeter layout) one possibility to explore is to employ a unique fiber type where both the Cherenkov and scintillation contributions are visible and well separated in the generated signal.

\section{Acknowledgements}
This work has been supported by the US Department of Energy, Office of Science (DE-SC001592), the Akdeniz University Scientific Research Projects Coordination Department (FDK-2017-2461), the Scientific and Technological Research Council of Turkey, T\"UBITAK, (No. 1059B141601412) and the Swiss National Science Foundation (FLARE grant Nr. 20FL20\texttt{\_}173602). The project has also been supported by European Union's Horizon 2020 research and innovation program under the grant agreement N$^\circ$654168 (AIDA 2020). We express our thanks to E. Auffray-Hillemanns for generously sharing the beam time at CERN. Jim Clarkin and Teo Tichindelean from Polymicro Technologies are thanked for their expertise, continued interest, and commitment to this R\&D project. Osamu Shinji from Kuraray is gratefully acknowledged for precious input and advice.

\bibliographystyle{JHEP}
\bibliography{3rdPaper}
\end{document}